\PassOptionsToPackage{unicode}{hyperref}
\PassOptionsToPackage{hyphens}{url}
\PassOptionsToPackage{dvipsnames,svgnames,x11names}{xcolor}
\documentclass[
  12pt]{article}

\usepackage{amsmath,amssymb}
\usepackage{iftex}
\ifPDFTeX
  \usepackage[T1]{fontenc}
  \usepackage[utf8]{inputenc}
  \usepackage{textcomp} 
\else 
  \usepackage{unicode-math}
  \defaultfontfeatures{Scale=MatchLowercase}
  \defaultfontfeatures[\rmfamily]{Ligatures=TeX,Scale=1}
\fi
\usepackage{lmodern}
\ifPDFTeX\else  
\fi
\IfFileExists{upquote.sty}{\usepackage{upquote}}{}
\IfFileExists{microtype.sty}{
  \usepackage[]{microtype}
  \UseMicrotypeSet[protrusion]{basicmath} 
}{}
\makeatletter
\@ifundefined{KOMAClassName}{
  \IfFileExists{parskip.sty}{%
    \usepackage{parskip}
  }{
    \setlength{\parindent}{0pt}
    \setlength{\parskip}{6pt plus 2pt minus 1pt}}
}{
  \KOMAoptions{parskip=half}}
\makeatother
\usepackage{xcolor}
\makeatletter
\ifx\paragraph\undefined\else
  \let\oldparagraph\paragraph
  \renewcommand{\paragraph}{
    \@ifstar
      \xxxParagraphStar
      \xxxParagraphNoStar
  }
  \newcommand{\xxxParagraphStar}[1]{\oldparagraph*{#1}\mbox{}}
  \newcommand{\xxxParagraphNoStar}[1]{\oldparagraph{#1}\mbox{}}
\fi
\ifx\subparagraph\undefined\else
  \let\oldsubparagraph\subparagraph
  \renewcommand{\subparagraph}{
    \@ifstar
      \xxxSubParagraphStar
      \xxxSubParagraphNoStar
  }
  \newcommand{\xxxSubParagraphStar}[1]{\oldsubparagraph*{#1}\mbox{}}
  \newcommand{\xxxSubParagraphNoStar}[1]{\oldsubparagraph{#1}\mbox{}}
\fi
\makeatother

\usepackage{breqn}
\usepackage{setspace}
\usepackage{float}
\usepackage{enumitem}
\usepackage{pgf}
\usepackage{subfigure}
\usepackage{pgfplots,pgfplotstable}
\pgfplotsset{compat=1.18} 
\usepackage{tikz}
\usetikzlibrary{patterns,decorations.pathreplacing}
\usepackage[flushleft]{threeparttable}

\usepackage{longtable,booktabs,array}
\usepackage{calc} 
\usepackage{etoolbox}
\makeatletter
\patchcmd\longtable{\par}{\if@noskipsec\mbox{}\fi\par}{}{}
\makeatother
\IfFileExists{footnotehyper.sty}{\usepackage{footnotehyper}}{\usepackage{footnote}}
\makesavenoteenv{longtable}
\usepackage{graphicx}
\makeatletter
\def\maxwidth{\ifdim\Gin@nat@width>\linewidth\linewidth\else\Gin@nat@width\fi}
\def\maxheight{\ifdim\Gin@nat@height>\textheight\textheight\else\Gin@nat@height\fi}
\makeatother
\setkeys{Gin}{width=\maxwidth,height=\maxheight,keepaspectratio}
\makeatletter
\def\fps@figure{htbp}
\makeatother

\makeatletter
\@ifpackageloaded{caption}{}{\usepackage{caption}}
\AtBeginDocument{%
\ifdefined\contentsname
  \renewcommand*\contentsname{Table of contents}
\else
  \newcommand\contentsname{Table of contents}
\fi
\ifdefined\listfigurename
  \renewcommand*\listfigurename{List of Figures}
\else
  \newcommand\listfigurename{List of Figures}
\fi
\ifdefined\listtablename
  \renewcommand*\listtablename{List of Tables}
\else
  \newcommand\listtablename{List of Tables}
\fi
\ifdefined\figurename
  \renewcommand*\figurename{Figure}
\else
  \newcommand\figurename{Figure}
\fi
\ifdefined\tablename
  \renewcommand*\tablename{Table}
\else
  \newcommand\tablename{Table}
\fi
}
\@ifpackageloaded{float}{}{\usepackage{float}}
\floatstyle{ruled}
\@ifundefined{c@chapter}{\newfloat{codelisting}{h}{lop}}{\newfloat{codelisting}{h}{lop}[chapter]}
\floatname{codelisting}{Listing}

\makeatother
\makeatletter
\@ifpackageloaded{caption}{}{\usepackage{caption}}
\@ifpackageloaded{subcaption}{}{\usepackage{subcaption}}
\makeatother

\ifLuaTeX
  \usepackage{selnolig}  
\fi
\usepackage[]{natbib}
\usepackage{bookmark}

\IfFileExists{xurl.sty}{\usepackage{xurl}}{} 
\hypersetup{
  pdftitle={IICI},
  pdfauthor={Anonymized},
  pdfkeywords={Restricted Parameter Space, Inequality Constraint, Model Selection, Parameter on the Boundary, Uniformly Valid Inference},
  colorlinks=true,
  linkcolor={blue},
  filecolor={Maroon},
  citecolor={Blue},
  urlcolor={Blue},
  pdfcreator={LaTeX via pandoc}}

\newcommand{\anon}{1}
\newcommand{\R}{\mathbb{R}}
\newtheorem{assumption}{Assumption}
\newtheorem{example}{Example}
\newtheorem{theorem}{Theorem}

\begin{document}

\def\spacingset#1{\renewcommand{\baselinestretch}%
{#1}\small\normalsize} \spacingset{1}


\if1\anon
{
  \title{\bf A Simple and Adaptive Confidence Interval when Nuisance Parameters Satisfy an Inequality}
  \author{Gregory Fletcher Cox\thanks{1 Arts Link \#06-02 AS2 117570 Singapore. Email: \href{mailto:ecsgfc@nus.edu.sg}{ecsgfc@nus.edu.sg}. ORCID: \href{https://orcid.org/0009-0009-9827-1434}{0009-0009-9827-1434}.}\hspace{.2cm}\\
    Department of Economics, National University of Singapore}
  \maketitle
} \fi

\if0\anon
{
  \bigskip
  \bigskip
  \bigskip
  \begin{center}
    {\LARGE\bf Title}
\end{center}
  \medskip
} \fi

\bigskip
\begin{abstract}
Inequalities may appear in many models. 
They can be as simple as assuming a parameter is nonnegative, possibly a regression coefficient or a treatment effect. 
This paper focuses on models with one inequality and proposes an inequality-imposed confidence interval (IICI) that has particularly attractive features. 
The IICI is simple in that it does not require simulations or tuning parameters. 
Also, the IICI is adaptive to the slackness of the inequality, uniformly valid, and never longer than the usual confidence interval. 
We demonstrate the IICI in two empirical applications. 
The first empirical application considers a regression when a coefficient is known to be nonpositive. 
The second empirical application considers an instrumental variables regression when the endogeneity of a regressor is known to be nonnegative. 
\end{abstract}

\noindent%
\vfill

\newpage
\spacingset{1.8} 

\section{Introduction}

Inequalities may appear in many models. 
In a linear regression, an inequality could be the assumption that the coefficient on a control variable is nonnegative. 
In an instrumental variables (IV) regression, an inequality could be the assumption that the sign of the endogeneity of a regressor is nonnegative. 
Other models that commonly have inequalities include random coefficient models, models of conditional heteroskedasticity, and structural equation models.\footnote{For random coefficient models with inequalities, see 
\cite{Ketz2018, Ketz2019} and the references therein. 
For models of conditional heteroskedasticity with inequalities, see \cite{FrancqZakoian2007, FrancqZakoian2009, FrancqZakoian2019}, \cite{Pedersen2017}, \cite{FrancqThieu2018}, and \cite{CavaliereNielsenRahbek2017, CNPR2022, CavalierePereraRahbek2024} and the references therein. 
For structural equation models with inequalities, see \cite{Stoeletal2006} and the references therein.} 
One can think of the inequalities as defining the boundary of the parameter space. 

A substantial literature documents and addresses the complications caused by inequalities on nuisance parameters in otherwise regular models. 
(By ``otherwise regular,'' we mean that a consistent and asymptotically normal estimator for the parameters exists. 
The only irregularity is the presence of the inequalities.) 
These complications arise when the true value of the parameters is at or near the boundary of the inequalities. 
Many applied researchers ignore the inequalities (or assume the true value of the parameters is in the interior of the parameter space) in order to avoid these complications. 
This paper focuses on the case that there is only one inequality and proposes a two-sided confidence interval (CI) that is particularly attractive. 
The case of one inequality is sufficiently common in empirical applications to justify special attention.\footnote{A rudimentary extension of the CI in this paper to multiple inequalities is to take an a priori convex combination of the inequalities. Other straightforward extensions appear to be invalid. We leave to future research the exploration of more sophisticated extensions.} 

A researcher that ignores the inequality calculates the usual CI (UCI) by adding and subtracting the standard error times the $1-\alpha/2$ standard normal quantile. 
The new CI, called the inequality-imposed CI (IICI), is equal to the UCI when the inequality is estimated to be sufficiently slack. 
This is an attractive feature of the IICI because it means the researcher does not need to report a different CI in order to be inequality-robust. 

If the inequality is assumed to hold with equality, then a researcher can calculate an equality-imposed CI (EICI) for the parameter of interest that uses the usual formula with a (weakly) smaller standard error. 
When the inequality is estimated to be sufficiently violated, then the IICI is equal to the EICI. 
This is an attractive feature of the IICI because it means the researcher can ignore the possible slackness of the inequality and use the CI that assumes equality. 

In between, the IICI simply interpolates between the UCI and the EICI. 
One of the endpoints of the IICI is equal to an endpoint of the UCI, while the other endpoint is equal to an endpoint of the EICI; see Figure \ref{Comparison1} for a picture. 
This is an attractive feature of the IICI because it does not depend on any tuning parameters or simulations. 
The endpoints of the IICI are a closed-form function of an asymptotically normal estimator and its asymptotic covariance matrix. 
This also means the IICI is never longer than the UCI. 
The length of the IICI shrinks monotonically as the inequality is estimated to be more violated until it reaches the length of the EICI. 
This is an attractive feature of the IICI because it means there is essentially no cost in terms of length to imposing an inequality if one is available. 

The final and most surprising feature of the IICI is that it is valid. 
Below, we show that the IICI is finite-sample valid in a normal model with known covariance matrix and asymptotically uniformly valid in an asymptotically normal model with consistently estimated asymptotic covariance matrix. 
The IICI also has exact coverage when the inequality holds with equality. 

\textbf{Related Literature.} 
It is well-known that estimators and standard test statistics have nonstandard limit theory when the true value of the parameters is at or near the boundary of the parameter space; see \cite{Chernoff1954}, \cite{Geyer1994}, \cite{Andrews1999, Andrews2001}, \cite{MoonSchorfheide2009}, and \cite{FanShi2023}. 
There are a couple of distinct problems where the complications due to a parameter on the boundary arise. 

(1) The first problem is inference on a parameter of interest with one or more inequalities on the nuisance parameters. 
(This is the problem considered in this paper, albeit with only one inequality.) 
In this problem, one is not trying to test whether the inequalities hold. 
Instead, one assumes the inequalities hold and uses them as ``overidentifying'' restrictions on the parameter of interest. 
\cite{MoonSchorfheide2009} show how this problem covers the general setting of moment inequality models when moment equalities are sufficient for identification. 
Several papers in the literature cover this problem. 
Some papers modify the estimator/test statistic so the asymptotic distribution continues to be normal or chi-squared even when the parameter is at or near the boundary; see \cite{CalzolariFiorentiniSentana2004}, \cite{Ketz2018}, and \cite{FrazierRenault2020}. 
This results in a CI that is asymptotically equivalent to the UCI. 
Some papers propose modifications to make tests inequality robust; see \cite{MoonSchorfheide2009}, \cite{Pedersen2017}, \cite{PedersenRahbek2019}, \cite{CNPR2022, CavalierePereraRahbek2024}, and \cite{FanShi2023}. 
Some papers use the inequalities as a source of information and propose hypothesis tests or CIs that have higher power or shorter length; see \cite{ElliottMullerWatson2015}, \cite{Ketz2018}, and \cite{KetzMcCloskey2023}. 
\cite{AndrewsGuggenberger2010ET} consider the one-inequality case and show using simulations that subsampling is not inequality-robust and that the two-sided CI with the inequality-imposed t-statistic is.\footnote{See Section \ref{Simulations} for the definition of the inequality-imposed t-statistic.} 
In Section \ref{Simulations}, we compare the IICI to CIs from these papers.\footnote{Some of these papers recommend hypothesis tests instead of CIs. Throughout this paper, when we refer to a CI from a particular paper that focuses on hypothesis testing, it is understood to be the CI that is calculated by collecting the hypotheses that fail to reject.} 

(2) A closely related but distinct problem is testing inequalities. 
This problem tests the hypothesis that the inequalities hold against the unrestricted alternative that at least one inequality does not. 
There is a large classical literature that covers this problem; see \cite{SilvapulleSen2005} and the references therein. 
Most of this literature is based on the (least-favorable) distribution of the likelihood-ratio statistic as a mixture of chi-squared distributions. 
\cite{SelfLiang1987} and \cite{KopylevSinha2011} point out that when there are nuisance parameters on the boundary, then the distribution of the likelihood-ratio statistic no longer takes this form. 
The literature on testing moment inequalities, including 
\cite{AndrewsSoares2010}, \cite{AndrewsBarwick2012}, \cite{RomanoShaikhWolf2014}, \cite{ChernozhukovChetverikovKato2019}, \cite{CoxShi2023}, and \cite{AndrewsRothPakes2023}, 
while focused on partially identified models, can also be applied to the testing inequalities problem. 

(3) A third problem arises when only the parameter of interest is subject to inequalities. 
That is, there are no nuisance parameters, or the nuisance parameters are not subject to the inequalities. 
Papers that cover this problem include \cite{Andrews2000}, \cite{FrancqZakoian2009}, \cite{AndrewsGuggenberger2010JoE}, \cite{MullerNorets2016}, \cite{CavaliereNielsenRahbek2017}, and \cite{FrancqThieu2018}. 
This problem is different because, under the null, the slackness of the inequalities is known. 

Another related literature consists of papers that discuss inference after model selection and adaptation to submodels. 
Papers that point out the challenges of inference after model selection and propose solutions include \cite{LeebPotscher2005}, \cite{AndrewsGuggenberger2009ECMA}, \cite{McCloskey2017, McCloskey2020}, and \cite{MuralidharanRomeroWuthrich2023}. 
For a class of regression models, \cite{ArmstrongKolesar2018} state a general bound on adapting to a submodel that does not apply in this case because it requires the parameter space to be centro-symmetric. 

\textbf{Empirical Applications.} There are several categories of commonly used applications that have one inequality on the parameters: 
(a) In an experiment with multiple treatments, one of the treatment effects may have known sign (nonnegative or nonpositive). 
(b) In an instrumental variables regression with one endogenous variable, it is common for the direction of the endogeneity (the covariance between the regressor and the error) to be known. 
(c) When estimating an Engel curve, the coefficient on income is nonnegative if the good is normal. 
(d) In discrete choice models that specify random utilities linearly as a function of price, the coefficient on price is typically assumed to be nonpositive. 
(e) When estimating a firm investment equation, the coefficient on Tobin's q is nonnegative if the firm faces convex adjustment costs of investment. 
(f) In production function estimation, the sum of the coefficients on the inputs is less than or equal to one under the assumption of a Cobb-Douglas production function with nonincreasing returns to scale. 
Any one of these categories has many applications that fit the setting of this paper. 

The first application revisits a linear regression in \cite{BlattmanJamisonSheridan2017}, hereafter BJS, that analyzes an experiment designed to evaluate the effects of cognitive behavioral therapy on crime and violence. 
BJS analyze the effects of two treatments together with their combination. 
The first treatment is the therapy, and the second treatment is an unconditional cash grant. 
The IICI can be calculated for the coefficients in the linear regression assuming the treatment effect for the cash grant is nonpositive (not increasing crime/violence). 
While the UCI for the treatment effect for therapy easily includes zero, we find that the IICI for the treatment effect for therapy is shorter and barely includes zero. 
Also, the IICI for the treatment effect for both treatments combined is shorter and more negative than the UCI. 

The second application revisits an IV regression in \cite{Shapiro2021}, hereafter S21, that analyzes an implicit subsidy given to industries with higher carbon emissions because they face lower tariffs and non-tariff barriers (NTBs). 
S21 uses an instrument to account for the possibility of measurement error in the measure of carbon emissions for an industry. 
The sign of the endogeneity of the measure of carbon emissions can be evaluated when the endogeneity is caused by measurement error. 
We find that for NTBs, the IICI is a subset of the UCI that excludes the smallest values of the carbon subsidy. 
This example demonstrates that the inequality need not be estimated to be violated in order for the IICI to be shorter than the UCI. 

\textbf{Outline.} 
Section \ref{Setting} presents the setup and examples. 
Section \ref{IICI_Definition} defines the IICI. 
Section \ref{Validity_Section} states theorems for validity of the IICI. 
Section \ref{Simulations} compares the IICI to other CIs in the literature by simulation. 
Section \ref{Applications} presents the empirical applications. 
Section \ref{Conclusion} concludes. 
An online Appendix contains proofs and additional comparisons. 
Code that implements the IICI in Stata and Matlab is available on the author's website. 

\section{Setting and Examples}
\label{Setting}

We consider a general setting with one inequality on a vector of parameters. 
Let $\theta$ denote a $k$-dimensional vector of parameters. 
The inequality on $\theta$ is given by $g(\theta)\le 0$, where $g(\cdot)$ is a known function of $\theta$. 
Let $\widehat\theta$ be an estimator of $\theta$. 
Below, we assume $\widehat\theta$ is consistent and asymptotically normal at the $\sqrt{n}$ rate, where $n$ is the sample size. 
We can think of $\widehat\theta$ as an unrestricted estimator; $\widehat\theta$ is not required to satisfy the inequalities. 
The existence of such estimators is a standard result in many identified parametric and semiparametric models. 
Such estimators may be defined by linear or IV regression, generalized method of moments (GMM), (quasi-) maximum likelihood, minimum distance, or indirect inference.\footnote{Sometimes, an objective function cannot be defined when the inequality is violated. Even in that case, \cite{Ketz2018} shows how to define a consistent and asymptotically normal estimator.} 
Let $\widehat V$ be a symmetric estimator of the asymptotic variance of $\sqrt{n}(\widehat\theta-\theta)$. 
Suppose the parameter of interest is the first element of $\theta$, $e'_1\theta$, where $e_j$ denotes the $j^{\text{th}}$ standard normal basis vector in $\R^k$ for $j\in\{2,...,k\}$. 
We give two examples of models that fit this setting. 

\begin{example}[Linear Regression]\label{ExLR}
Consider a linear regression model: 
\begin{equation}
Y_i=\beta'X_i+\epsilon_i, 
\end{equation}
where $Y_i$ is the dependent variable, $\beta$ is $k$-vector of coefficients, $X_i$ is a $k$-vector of regressors, $\epsilon_i$ is an unobserved error, and $i\in\{1,...,n\}$. 
Suppose $\beta_1=e'_1\beta$ is a scalar parameter of interest. 
An inequality could be the restriction that one of the other coefficients is nonnegative or nonpositive. 
In that case, $g(\beta)=\pm e'_j\beta$. 
The setting is satisfied if we take $\widehat\beta$ to be the least squares estimator and $\widehat V$ to be an estimator of the asymptotic variance of $\widehat\beta$. 

The assumption that a coefficient is nonnegative/nonpositive is reasonable in many applications of linear regressions. 
Here, we describe an example that evaluates multiple treatments. 
Suppose $X_{1i}=e'_1X_i$ and $X_{2i}=e'_2 X_i$ indicate treatments from a randomized controlled trial with two disjoint treatment groups. 
Then, $\beta_1$ and $\beta_2$ are the average treatment effects.\footnote{More generally, it is important that interactions between the treatment indicators are included in the regression when treatment groups are not disjoint; see \cite{MuralidharanRomeroWuthrich2023}.} 
\cite{KetzMcCloskey2023} point out that many treatments can reasonably be assumed to have nonnegative average effects. 
Indeed, an ethics board may not approve a proposed treatment that has a negative average effect on the participants. 
If the second treatment has a nonnegative average effect, then $\beta_2\ge 0$ and the IICI can be implemented for $\beta_1$. 
In Section \ref{Application1}, we implement the IICI in an empirical application with this setting. 

We can try to assess the effect of imposing the inequality in a simple case with multiple treatments. 
Suppose there are only two disjoint treatment groups, $T_1$ and $T_2$, and a control group, $C$. 
Let $\overline{Y}_S$ denote the average outcome over individuals in $S$ for $S\subset \{1,...,n\}$. 
Then, the least-squares estimators of $\beta_1$ and $\beta_2$ are $\widehat\beta_1=\overline{Y}_{T_1}-\overline{Y}_{C}$ and $\widehat\beta_2=\overline{Y}_{T_2}-\overline{Y}_C$. 
If $\beta_2$ is assumed to be nonnegative but $\widehat\beta_2<0$, then there is clearly some error in the estimate of $\beta_2$. 
$\widehat \beta_2=\overline{Y}_{T_2}-\overline{Y}_C$ must be under-estimating the true value of $\beta_2$. 
Either $\overline{Y}_{T_2}$ under-estimates the mean of $Y_i$ over $i\in T_2$ or $\overline{Y}_C$ over-estimates the mean of $Y_i$ over $i\in C$ (or both). 
This error can be partially corrected by imposing the inequality. 
Let $\ddot\beta_1$ and $\ddot\beta_2$ denote the least-squares estimators for $\beta_1$ and $\beta_2$ that impose the inequality. 
If $\widehat\beta_2<0$, then $\ddot\beta_2=0$ and $\ddot\beta_1=\overline{Y}_{T_1}-\overline{Y}_{C\cup T_2}$. 
This implicitly estimates the mean of $Y_i$ over $i\in C$ using the average value of $Y_i$ over $i\in C\cup T_2$. 
Essentially, the individuals in $T_2$ are counted as part of the control group. 
Note that $\overline{Y}_{C\cup T_2}<\overline{Y}_{C}$. 
This means that imposing the inequality increases the estimate of $\beta_1$. 
This is intuitive because the original estimator used $\overline{Y}_C$, which is likely to over-estimate the mean of $Y_i$ over $i\in C$ when $\widehat\beta_2<0$. 
Also, $\overline{Y}_{C\cup T_2}$ is estimated with a larger sample, resulting in a smaller standard error. 
These effects of imposing the inequality provide some intuition for why and how the IICI, defined in Section \ref{IICI_Definition}, differs from the UCI. 
\end{example}

\begin{example}[Instrumental Variables Regression]\label{ExIV}
Consider an IV regression: 
\begin{align}
Y_i&=\beta'X_i+\delta'W_i+\epsilon_i\\
X_i&=\Pi_1Z_i+\Pi_2W_i+u_i,  
\end{align}
where $Y_i$ is the dependent variable, $X_i$ is a $k_1$-vector of endogenous regressors, $W_i$ is a $k_2$-vector of exogenous regressors, $\beta$ and $\delta$ are vectors of coefficients, $Z_i$ is an $\ell$-vector of excluded exogenous instruments, $\Pi_1$ and $\Pi_2$ are matrices of coefficients, $\epsilon_i$ and $u_i$ are unobserved errors, and $i\in\{1,...,n\}$. 
The endogeneity of the endogenous regressors is determined by $\gamma=\mathbb{E}(X_i\epsilon_i)$.\footnote{The regressors in $X_i$ are called endogenous because we do not \textit{assume} the endogeneity is zero. The true values of the components of $\gamma$ can be zero or nonzero.} 

Knowledge of the sign of a component of $\gamma$ can be exploited as an inequality on a nuisance parameter.\footnote{A known sign of one of the coefficients, $\beta$ or $\delta$, also counts as an inequality. In this case, the usual IV/GMM estimator and asymptotic variance can be used to calculate the IICI. Relatedly, a known sign of a first-stage coefficient in $\Pi_1$ or $\Pi_2$, as considered in \cite{AndrewsArmstrong2017}, also counts, although the moments used to estimate $\Pi_1$ and $\Pi_2$ need to be added to the GMM specification. Also note that the model must be identified. This excludes the case of one endogenous regressor and one instrument with a known sign of the first-stage coefficient because the first-stage coefficient must be nonzero for strong identification.} 
It is common for empirical studies to suggest a sign of $\gamma$ when explaining why an instrument is needed. 
\cite{MoonSchorfheide2009} review empirical studies and report: ``In almost all of the papers the authors explicitly stated their beliefs about the sign of the correlation between the endogenous regressor and the error term; yet none of the authors exploited the resulting inequality moment condition in their estimation.'' 
Specifying the sign of $\gamma$ is the first step in signing the omitted variable bias of the least-squares estimator. 
A special case of this is when $X_i$ includes classical measurement error. 
Suppose $e'_j X_i$ is measured with error for $j\in\{1,...,k\}$. 
Then, $e'_j\gamma=-e'_j\beta\sigma^2$, where $\sigma^2$ denotes the variance of the measurement error. 
Then, the sign of $e'_j\gamma$ is the negative of the sign of $e'_j\beta$. 
In Section \ref{Application2}, we implement the IICI in an empirical application with this setting. 

We can estimate $\beta$ and $\delta$ jointly with $\gamma$ by GMM. 
Suppose $\Pi_1$ has rank $k_1$ so the structural coefficients are strongly identified. 
Then, $\beta$ and $\delta$ are identified using the moments $\mathbb{E}Z_i(Y_i-\beta'X_i-\delta'W_i)=0$ and $\mathbb{E}W_i(Y_i-\beta'X_i-\delta'W_i)=0$. 
We identify $\gamma$ by including the moments $\mathbb{E}X_i(Y_i-\beta'X_i-\delta'W_i)-\gamma=0$. 
The GMM estimator is asymptotically normal with the usual asymptotic variance-covariance matrix. 
We point out that the GMM estimator for the structural coefficients (and their standard errors) is unchanged by the inclusion of the moments for $\gamma$. 
Code that demonstrates how to jointly estimate $\beta$, $\delta$, and $\gamma$ in Stata is available on the author's website. 
\end{example}

\section{An Inequality-Imposed Confidence Interval}
\label{IICI_Definition}

The UCI is defined to be 
\begin{equation}
\text{UCI}=[e'_1\widehat\theta-\hat s z_{1-\alpha/2}, e'_1\widehat\theta+\hat s z_{1-\alpha/2}], 
\end{equation}
where $\hat s=n^{-1/2}(e'_1\widehat Ve_1)^{1/2}$ is the standard error of $e'_1\widehat\theta$ and $z_{1-\alpha/2}$ denotes the $1-\alpha/2$ standard normal quantile. 
The UCI is appealing because it does not require any modification to standard practice. 
The downside is that it does not use the inequality. 

We next define the EICI. 
Suppose, for simplicity, $g(\theta)$ is linear in $\theta$ with $g(\theta)=a'\theta+b$, where $a$ and $b$ are constants. 
When the inequality is assumed to hold with equality, the resulting equality-imposed estimator (EIE) is given by 
\begin{equation}
\ddot\theta=\widehat\theta-\widehat Va(a'\widehat Va)^{-1}g(\widehat\theta). \label{EIE_new}
\end{equation}
The standard error for the EIE is given by 
\begin{equation}
\ddot s = n^{-1/2}\left(e'_1\widehat Ve_1-e'_1\widehat V a (a' \widehat V a)^{-1} a'\widehat V e_1\right)^{1/2}. \label{EISE}
\end{equation}
This is (weakly) smaller than $\hat s$ with equality if $e'_1\widehat V a=0$. 
The (weakly) smaller standard error represents the fact that imposing the equality on the estimator may reduce the noise in estimating $e'_1\theta$. 
The EICI is then given by 
\begin{equation}
\text{EICI}=[e'_1\ddot\theta-\ddot s z_{1-\alpha/2}, e'_1\ddot\theta+\ddot s z_{1-\alpha/2}]. \label{EICI}
\end{equation}
When $g(\theta)$ is not linear, then the EICI can be defined using (\ref{EIE_new})-(\ref{EICI}) with $a$ replaced by $\hat a=\frac{d}{d\theta}g(\widehat\theta)$. 

The IICI is defined to transition between the UCI, when the inequality is estimated to be sufficiently slack, and the EICI, when the inequality is estimated to be sufficiently violated. 
Let 
\begin{equation}
\ddot c=\begin{cases}
(e'_1\widehat V a)^{-1}(a'\widehat Va)(\hat s-\ddot s)z_{1-\alpha/2} & \text{ if } e'_1\widehat V a\neq 0\\
0&\text{ if } e'_1\widehat V a=0
\end{cases}\label{ddotc_def}
\end{equation}
be a threshold that determines when to transition between the UCI and the EICI. 
Let $\text{IICI}=[\ell, u]$, where 
\begin{align}
\ell=\begin{cases}
e'_1\widehat\theta-\hat s z_{1-\alpha/2}&\text{ if }g(\widehat\theta)\le \ddot c\\
e'_1\ddot\theta-\ddot s z_{1-\alpha/2}&\text{ if }g(\widehat\theta)>\ddot c
\end{cases}\label{IICI_ell} \quad\text{ and }\quad
u=\begin{cases}
e'_1\widehat\theta+\hat s z_{1-\alpha/2}&\text{ if }g(\widehat\theta)\le -\ddot c\\
e'_1\ddot\theta+\ddot s z_{1-\alpha/2}&\text{ if }g(\widehat\theta)> -\ddot c
\end{cases}. 
\end{align}
When $e'_1\widehat V a=0$, then $e'_1\widehat\theta=e'_1\ddot\theta$ and $\hat s=\ddot s$, so $\text{IICI}=\text{UCI}=\text{EICI}$. 

\begin{figure}
\begin{center}
\scalebox{0.9}{
\subfigure{
\begin{tikzpicture}
\pgfmathsetmacro{\cov}{0.7}
\pgfmathsetmacro\horizontallength{6}
\pgfmathsetmacro\horizontalshift{0}
\def\verticallength{4.5}
\def\verticalshift{-0.5}
\pgfmathsetmacro{\EIEleft}{-\cov*(-\horizontallength+\horizontalshift)}
\pgfmathsetmacro{\EIEright}{-\cov*(\horizontallength+\horizontalshift)}
\begin{axis}[width=\textwidth, height=0.75\textwidth, ymin=-5,ymax=4,ylabel=$e'_1\theta$,xmin=-6,xmax=6,xlabel=$e'_2\widehat\theta$,legend style={at={(axis cs:(-5.4,-3)},anchor=north west}]
\draw[-] (-\horizontallength+\horizontalshift,0)--(\horizontallength+\horizontalshift,0); 
\draw[-] (0,-\verticallength+\verticalshift)--(0,\verticallength+\verticalshift); 
\draw[red,dotted,thick] (\horizontallength+\horizontalshift,\EIEright)--(-\horizontallength+\horizontalshift,\EIEleft); 
\draw[gray!50, dashed] (0.8,-5.4) -- (0.8,5.4);
\draw[gray!50, dashed] (-0.8,-5.4) -- (-0.8,5.4);
\addplot[blue, dashed,thick] table [x=theta2, y=upper, col sep=comma]{UCI_rho_70.csv};
\addplot[red, dashdotted,thick] table [x=theta2, y=upper, col sep=comma]{EICI_rho_70.csv};
\addplot[violet, -,thick] table [x=theta2, y=upper, col sep=comma]{IICI_rho_70.csv};
\addplot[blue, dashed,thick] table [x=theta2, y=lower, col sep=comma]{UCI_rho_70.csv};
\addplot[red, dashdotted,thick] table [x=theta2, y=lower, col sep=comma]{EICI_rho_70.csv};
\addplot[violet, -,thick] table [x=theta2, y=lower, col sep=comma]{IICI_rho_70.csv};
\addlegendentry{{\footnotesize UCI}};
\addlegendentry{{\footnotesize EICI}};
\addlegendentry{{\footnotesize IICI}};
\end{axis}
\draw (\horizontallength+0.5,2*\verticallength+2*\verticalshift+3) node {Panel A: Confidence Intervals};
\end{tikzpicture}
}
}

\scalebox{0.9}{
\subfigure{
\begin{tikzpicture}
\begin{axis}[width=0.5\textwidth, height=0.4\textwidth, ymin=0.8,ymax=1,xmin=-5,xmax=0,xlabel=$e'_2\theta$]
\addplot[blue, dashed,thick] table [x=theta2, y=CP, col sep=comma]{UCI_CP_AL_rho_70.csv};
\addplot[red, dashdotted,thick] table [x=theta2, y=CP, col sep=comma]{EICI_CP_AL_rho_70.csv};
\addplot[violet, -, thick] table [x=theta2, y=CP, col sep=comma]{IICI_CP_AL_rho_70.csv};
\end{axis}
\draw (3,5.3) node {Panel B: Coverage Probability};
\end{tikzpicture}
}
}
\scalebox{0.9}{
\subfigure{
\begin{tikzpicture}
\begin{axis}[width=0.5\textwidth, height=0.4\textwidth, ymin=2,ymax=4,xmin=-5,xmax=0,xlabel=$e'_2\theta$]
\addplot[blue, dashed,thick] table [x=theta2, y=AL, col sep=comma]{UCI_CP_AL_rho_70.csv};
\addplot[red, dashdotted,thick] table [x=theta2, y=AL, col sep=comma]{EICI_CP_AL_rho_70.csv};
\addplot[violet, -, thick] table [x=theta2, y=AL, col sep=comma]{IICI_CP_AL_rho_70.csv};
\end{axis}
\draw (3,5.3) node {Panel C: Average Length};
\end{tikzpicture}
}
}
\end{center}
\vspace{-3mm}
\caption{Comparing the UCI, EICI, IICI, and EIE.\vspace{2mm}\\Notes: Panel A depicts the UCI, EICI, and IICI for $e'_1\theta$ when $e'_1\widehat\theta=0$ as functions of $e'_2\widehat\theta$, with $g(\theta)=e'_2\theta$ and $\text{corr}(e'_1\widehat\theta,e'_2\widehat\theta)=0.7$. 
The blue dashed lines depict the upper and lower bounds for the UCI. 
The red dotted line depicts the EIE. 
The red dash-dotted lines depict the upper and lower bounds for the EICI. 
The purple solid lines depict the upper and lower bounds for the IICI. 
The kink occurs at $e'_1\widehat\theta=\pm \ddot c\approx \pm 0.8$. 
Panels B and C depict the coverage probabilities and average lengths of the UCI, EICI, and IICI as functions of $e'_2\theta$, calculated with $10^5$ simulations. 
}
\label{Comparison1}
\end{figure}
Figure \ref{Comparison1} illustrates the UCI, EICI, and IICI in the case that $\theta$ is two-dimensional, $g(\theta)=e'_2\theta$ is assumed to be nonpositive, and $\widehat V=n\left[\begin{smallmatrix}1&0.7\\0.7&1\end{smallmatrix}\right]$. 
(Scaling $\widehat V$ by $n$ removes the dependence of the CIs on $n$ and, in particular, prevents their length from shrinking to zero at the $n^{-1/2}$ rate.) 
Figure \ref{Comparison1} depicts the CIs for $e'_1\theta$ as a function of $e'_2\widehat\theta$, which is the estimated slackness/violation of the inequality. 
The value of $e'_1\widehat\theta$ is normalized to $0$. 
The CIs are the vertical intervals between the upper and lower bounds. 
We make some remarks on Figure \ref{Comparison1}. 

\textbf{Remarks.} (1) In Panel A, the endpoints of the UCI (EICI) are depicted by dashed blue (dash-dotted red) lines. 
The endpoints of the IICI are depicted by the solid purple lines. 
When $e'_2\widehat\theta<-\ddot c$, the inequality is estimated to be sufficiently slack, and $IICI=UCI$. 
When $e'_2\widehat\theta\in[-\ddot c, \ddot c]$, the upper bound of the IICI is equal to the upper bound of the EICI and the lower bound of the IICI is equal to the lower bound of the UCI. 
When $e'_2\widehat\theta>\ddot c$, the inequality is estimated to be sufficiently violated, and $IICI=EICI$. 

(2) In Panel A, the EIE is depicted by the dotted red line. 
The EIE is negative for values of $e'_2\widehat\theta>0$. 
This makes sense because $e'_1\widehat\theta$ and $e'_2\widehat\theta$ are positively correlated. 
When $e'_2\widehat\theta>0$, then $e'_2\theta$ must be over-estimated. 
This indicates that $e'_1\widehat\theta=0$ also over-estimates $e'_1\theta$, and so the EIE is negative. 
This explains why the EIE has a negative slope as a function of $e'_2\widehat\theta$. 

(3) Panel B depicts the coverage probabilities of the UCI, EICI, and IICI as functions of $e'_2\theta$. 
The UCI has $95\%$ coverage probability for any value of $e'_2\theta$. 
The EICI only has $95\%$ coverage probability for $e'_2\theta=0$. 
The EICI is not designed to cover $e'_1\theta$ when the inequality is slack. 
The IICI has coverage probability greater than $95\%$ for any value of $e'_2\theta< 0$, with equality when $e'_2\theta=0$ and in the limit as $e'_2\theta\rightarrow -\infty$. 

(4) Panel C depicts the average lengths of the UCI, EICI, and IICI as functions of $e'_2\theta$. 
The UCI always has length $2\times z_{1-\alpha/2}\times \hat s\approx 3.92$ (because $\hat s=1$). 
The EICI always has length $2\times z_{1-\alpha/2}\times\ddot s\approx 2.78$ (because $\ddot s\approx 0.71$). 
The average length of the IICI is equal to the length of the UCI in the limit as $e'_2\theta\rightarrow-\infty$ and decreases as a function of $e'_2\theta$. 
When $e'_2\theta=0$, the average length of the IICI is exactly halfway between the length of the UCI and the length of the EICI.  

(5) If $e'_2\widehat\theta$ is very large, then the inequality is estimated to be very violated. 
In that case, it is possible the IICI is disjoint from the UCI. 
That is, imposing the inequality gives a completely distinct interval from ignoring the inequality. 
This is very unlikely if the inequality holds. 
In this case, the one-sided test for $H_0: e'_2\theta\le 0$ would reject if $e'_2\widehat\theta>z_{1-\alpha}\approx 1.64$. 

\section{Validity of the IICI}
\label{Validity_Section}

Most of the attractive properties that have been claimed to hold for the IICI follow directly from the definition. 
The surprising property of the IICI is its validity. 
We prove validity in two cases. 
The first case covers the finite-sample model with normally distributed $\widehat\theta$, known covariance matrix, and a linear inequality. 
The second case covers asymptotically normal $\widehat\theta$, consistently estimated asymptotic covariance matrix, and a possibly nonlinear inequality. 

\subsection{Finite-Sample Validity}

Let $\theta_0\in\R^k$ denote the true value. 
The following assumption specifies the finite-sample model. 

\begin{assumption}\label{Normal_IICI}
\begin{enumerate}[label=(\alph*)]
\item $\sqrt{n}(\widehat\theta-\theta_0)\sim N(0,V)$ for some covariance matrix, $V$. 
\item $V$ is positive definite, and $\widehat V=V$ almost surely. 
\item $g(\theta)$ is linear in $\theta$. Let $g(\theta)=a'\theta+b$ for some $a\in\R^k$ and $b\in\R$. 
\item $e_1e'_1a\neq a$. 
\end{enumerate}
\end{assumption}
\noindent\textbf{Remark.} Part (a) specifies the multivariate normal distribution for the estimator. 
Part (b) requires the covariance matrix to be positive definite and known. 
Part (c) focuses on a linear inequality. 
Part (d) requires $a$ to not be a scalar multiple of $e_1$. 
This ensures the inequality is (at least partly) a function of the nuisance parameters. 
The parameter of interest can have a nonzero coefficient, but it cannot be the only parameter with a nonzero coefficient. 
This is somewhat more general than the title of this paper suggests.\footnote{The title of this paper suggests the inequality is only on the nuisance parameters, but actually we allow the inequality to be a function of all the parameters as long as it is not exclusively a function of the parameter of interest.} 

Let $P(\cdot)$ denote the probability calculated with respect to the distribution of $\widehat\theta$ specified in Assumption \ref{Normal_IICI}(a). 

\begin{theorem}\label{Thm_Normal_IICI}
Under Assumption \ref{Normal_IICI}, if $g(\theta_0)\le 0$, then $P(e'_1\theta_0\in \textnormal{IICI})\ge 1-\alpha$. 
Furthermore, if $g(\theta_0)=0$, then $P(e'_1\theta_0\in \textnormal{IICI})=1-\alpha$. 
\end{theorem}

\noindent\textbf{Remarks.} (1) Theorem \ref{Thm_Normal_IICI} establishes finite-sample validity of the IICI for any value of $\theta_0$ that satisfies the inequality. 
If $\theta_0$ satisfies the inequality with equality, then the IICI has exact coverage probability. 

(2) The proof of Theorem \ref{Thm_Normal_IICI} uses a new argument based on translating the normal distribution; see Lemma 3 in the online Appendix. 

\subsection{Asymptotic Uniform Validity}

We next show that the IICI is asymptotically uniformly valid in an asymptotically normal model. 
We state two theorems, one with linear $g(\cdot)$ and one with nonlinear $g(\cdot)$. 
Let $F\in\mathcal{F}_n$ specify the distribution of $\widehat\theta$ and $\widehat V$, where $\mathcal{F}_n$ is a parameter space that is allowed to depend on $n$. 
Also suppose the true value of $\theta$ depends on $F$, which we denote by $\theta_F$. 
The following assumption specifies the asymptotic model. 

\begin{assumption}\label{Asy_IICI2}
For every sequence $\{F_n\}_{n=1}^{\infty}$ such that $F_n\in\mathcal{F}_n$, and for every subsequence, $n_m$, there exists a further subsequence, $n_q$, and there exists a positive definite and symmetric matrix, $V$, such that: $\sqrt{n_q}(\widehat\theta-\theta_{F_{n_q}})\rightarrow_d N(0,V)$ and $\widehat V\rightarrow_p V$ as $q\rightarrow\infty$. 
\end{assumption}

\noindent\textbf{Remark.} Assumption \ref{Asy_IICI2} is essentially the requirement that $\widehat\theta$ is asymptotically normal and $\widehat{V}$ is a consistent estimator of the asymptotic covariance matrix. 
Assumption \ref{Asy_IICI2} is stated using an extra subsequencing step that makes the convergence uniform over $F\in\mathcal{F}_n$. 
The extra subsequencing step is innocuous in practice because typical arguments for asymptotic normality, including central limit theorems that allow for dependence and/or triangular arrays, can also be applied along subsequences. 

Let $P_{F}(\cdot)$ denote the probability calculated with respect to the distribution of $\widehat\theta$ and $\widehat V$ specified by $F$. 

\begin{theorem} \label{Thm_Asy_linear}
Under Assumption \ref{Normal_IICI}(c,d) and Assumption \ref{Asy_IICI2}, 
\[
\liminf_{n\rightarrow\infty}\inf_{\{F\in\mathcal{F}_n: g(\theta_F)\le 0\}}P_{F}(e'_1\theta_F\in \textnormal{IICI})\ge 1-\alpha. 
\]
Furthermore, if $\{F_n\}_{n=1}^\infty$ is a sequence such that $F_n\in\mathcal{F}_n$ and $\sqrt{n}g(\theta_{F_n})\rightarrow 0$, then $P_{F_n}(e'_1\theta_{F_n}\in \textnormal{IICI})\rightarrow1-\alpha$ as $n\rightarrow\infty$. 
\end{theorem}

\noindent\textbf{Remark.} Theorem \ref{Thm_Asy_linear} establishes asymptotic uniform validity of the IICI over values of $F$ for which $\theta_F$ satisfies the inequality. 
If $g(\theta_F)$ is sufficiently close to zero, then the IICI has exact asymptotic coverage probability. 

Theorem \ref{Thm_Asy_linear} requires $g(\theta)$ to be linear. 
We next generalize Theorem \ref{Thm_Asy_linear} to allow $g(\theta)$ to be nonlinear. 
The following assumption states restrictions on $g(\theta)$. 

\begin{assumption}\label{Asy_IICI}
\begin{enumerate}[label=(\alph*)]
\item There exists a set, $\mathcal{U}\subset\R^k$, such that $g:\mathcal{U}\rightarrow\R$ is differentiable with derivative $G: \mathcal{U}\rightarrow\R^k$ and such that $G(\theta)$ is uniformly continuous on $\mathcal{U}$. 
\item For every sequence, $\{F_n\}_{n=1}^{\infty}$, such that $F_n\in\mathcal{F}_n$, there exists an $\epsilon>0$ such that $B(\theta_{F_n},\epsilon)\subset\mathcal{U}$ eventually as $n\rightarrow\infty$, where $B(\theta,\epsilon)=\{x\in\R^k: \|x-\theta\|<\epsilon\}$. 
\item $0<\inf_{\theta\in\mathcal{U}}\|e_1e'_1G(\theta)-G(\theta)\|$ and $\sup_{\theta\in\mathcal{U}}\|G(\theta)\|<\infty$. 
\end{enumerate}
\end{assumption}

\noindent\textbf{Remarks.} (1) Part (a) requires $g(\theta)$ to be defined and differentiable on a set, $\mathcal{U}$, where the derivative is uniformly continuous.
Part (b) requires $\mathcal{U}$ to contain $\theta_F$ in its interior, where the interiority requirement is uniform over $F\in\mathcal{F}_n$. 
Part (c) requires $G(\theta)$ to not be a scalar multiple of $e_1$ and $G(\theta)$ to be bounded in a way that is uniform over $\theta\in\mathcal{U}$. 

(2) Assumption \ref{Asy_IICI} is trivially satisfied under Assumption \ref{Normal_IICI}(c,d) by setting $\mathcal{U}=\R^k$ because, in that case, $G(\theta)=a$ does not depend on $\theta$. 

(3) In nonlinear models, it is common to assume the parameter space for $\theta$, say $\Theta$, is compact. 
In that case, Assumption \ref{Asy_IICI} is satisfied by taking $\mathcal{U}$ to be a small expansion of $\Theta$, assuming $g(\theta)$ is continuously differentiable on $\mathcal{U}$, and assuming $e_1e'_1G(\theta)\neq G(\theta)$ for all $\theta\in\mathcal{U}$. 

\begin{theorem}\label{Thm_Asy_nonlinear}
Under Assumptions \ref{Asy_IICI2} and \ref{Asy_IICI}, 
\[
\liminf_{n\rightarrow\infty}\inf_{\{F\in\mathcal{F}_n: g(\theta_F)\le 0\}}P_{F}(e'_1\theta_F\in \textnormal{IICI})\ge 1-\alpha. 
\]
Furthermore, if $\{F_n\}_{n=1}^\infty$ is a sequence such that $F_n\in\mathcal{F}_n$ and $\sqrt{n}g(\theta_{F_n})\rightarrow 0$, then $P_{F_n}(e'_1\theta_{F_n}\in \textnormal{IICI})\rightarrow1-\alpha$ as $n\rightarrow\infty$. 
\end{theorem}

\noindent\textbf{Remarks.} (1) Theorem \ref{Thm_Asy_nonlinear} establishes asymptotic uniform validity of the IICI for nonlinear $g(\theta)$. 
As in Theorem \ref{Thm_Asy_linear}, if $g(\theta_F)$ is sufficiently close to zero, then the IICI has exact asymptotic coverage probability. 

(2) The proof of Theorem \ref{Thm_Asy_nonlinear} shows that the event $\{e'_1\theta_F\in \textnormal{IICI}\}$ converges to a limiting event based on the finite-sample normal model. 
The limiting coverage probability can then be evaluated using Theorem \ref{Thm_Normal_IICI}. 
The proof of Theorem \ref{Thm_Asy_linear} is omitted because it follows from Theorem \ref{Thm_Asy_nonlinear}, using the fact that Assumption \ref{Normal_IICI}(c,d) implies Assumption \ref{Asy_IICI} with $\mathcal{U}=\R^k$. 

\section{Comparison to Other Confidence Intervals}
\label{Simulations}

This section compares the IICI to other CIs available in the literature. 
We consider (1a) a likelihood-ratio CI (LRCI), calculated by inverting the likelihood-ratio test with a chi-squared critical value, 
(1b) a size-corrected version of the LRCI (SCLRCI) that simulates a fixed critical value, 
(2) a conditional likelihood-ratio CI (CLRCI), calculated by inverting the conditional likelihood-ratio (CLR) test proposed in \cite{Ketz2018}, 
(3) the CI defined using the inequality-imposed t-statistic (IITCI) considered in \cite{AndrewsGuggenberger2010ET}, 
(4) a CI calculated by inverting a test from \cite{ElliottMullerWatson2015}, called EMWCI, and 
(5) the simple and short CI (SSCI) recommended by \cite{KetzMcCloskey2023}. 

Overall, we see that the IICI has coverage probability above 95\% and competitive average length. 
These properties combine with the attractive features of the IICI (validity, simplicity, adaptation, never longer than the UCI, and monotonically decreasing length in the estimated violation of the inequality) to justify the recommendation to use the IICI in models with one inequality. 

Throughout this section, we use the example from Figure \ref{Comparison1}, where $\theta$ is two-dimensional, $g(\theta)=e'_2\theta$ is assumed to be nonpositive, $\widehat\theta$ is normally distributed, and $\widehat V=n\left[\begin{smallmatrix}1&\rho\\\rho&1\end{smallmatrix}\right]$ with $\rho=0.7$. 
Figures \ref{Comparison_rho_70_2} and \ref{Comparison_rho_70_3} depict these CIs, together with their simulated coverage probabilities and average lengths as a function of the slackness of the inequality. 
The online Appendix considers different values of $\rho$ and reports similar conclusions. 

\begin{figure}
\begin{center}
\scalebox{0.9}{
\subfigure{
\begin{tikzpicture}
\pgfmathsetmacro{\cov}{0.7}
\pgfmathsetmacro\horizontallength{6}
\def\verticallength{4.5}
\def\verticalshift{-0.5}
\pgfmathsetmacro\horizontalshift{0}
\pgfmathsetmacro{\EIEleft}{-\cov*(-\horizontallength+\horizontalshift)}
\pgfmathsetmacro{\EIEright}{-\cov*(\horizontallength+\horizontalshift)}
\begin{axis}[width=\textwidth, height=0.75\textwidth, ymin=-5,ymax=4,ylabel=$e'_1\theta$,xmin=-6,xmax=6,xlabel=$e'_2\widehat\theta$,legend style={at={(axis cs:(-5.4,-3)},anchor=north west}]
\draw[-] (-\horizontallength+\horizontalshift,0)--(\horizontallength+\horizontalshift,0); 
\draw[-] (0,-\verticallength+\verticalshift)--(0,\verticallength+\verticalshift); 
\addplot[blue, densely dotted, thick] table [x=theta2, y=upper, col sep=comma]{LR_rho_70.csv};
\addplot[red, loosely dashed, thick] table [x=theta2, y=upper, col sep=comma]{LRSC_rho_70.csv};
\addplot[teal, dashdotted,thick] table [x=theta2, y=upper, col sep=comma]{K18_rho_70.csv};
\addplot[violet, -, thick] table [x=theta2, y=upper, col sep=comma]{IICI_rho_70.csv};
\addplot[blue, densely dotted, thick] table [x=theta2, y=lower, col sep=comma]{LR_rho_70.csv};
\addplot[red, loosely dashed, thick] table [x=theta2, y=lower, col sep=comma]{LRSC_rho_70.csv};
\addplot[teal, dashdotted,thick] table [x=theta2, y=lower, col sep=comma]{K18_rho_70.csv};
\addplot[violet, -, thick] table [x=theta2, y=lower, col sep=comma]{IICI_rho_70.csv};
\addlegendentry{{\footnotesize LRCI}};
\addlegendentry{{\footnotesize SCLRCI}};
\addlegendentry{{\footnotesize CLRCI}};
\addlegendentry{{\footnotesize IICI}};
\end{axis}
\draw (\horizontallength+0.5,2*\verticallength+2*\verticalshift+3) node {Panel A: Confidence Intervals};
\end{tikzpicture}
}
}

\scalebox{0.9}{
\subfigure{
\begin{tikzpicture}
\begin{axis}[width=0.5\textwidth, height=0.4\textwidth, ymin=0.9,ymax=1,xmin=-5,xmax=0,xlabel=$e'_2\theta$]
\addplot[blue, densely dotted, thick] table [x=theta2, y=CP, col sep=comma]{LR_CP_AL_rho_70.csv};
\addplot[red, loosely dashed, thick] table [x=theta2, y=CP, col sep=comma]{LRSC_CP_AL_rho_70.csv};
\addplot[teal, dashdotted,thick] table [x=theta2, y=CP, col sep=comma]{K18_CP_AL_rho_70.csv};
\addplot[violet, -, thick] table [x=theta2, y=CP, col sep=comma]{IICI_CP_AL_rho_70.csv};
\end{axis}
\draw (3,5.3) node {Panel B: Coverage Probability};
\end{tikzpicture}
}
}
\scalebox{0.9}{
\subfigure{
\begin{tikzpicture}
\begin{axis}[width=0.5\textwidth, height=0.4\textwidth, ymin=3.2,ymax=4.2,xmin=-5,xmax=0,xlabel=$e'_2\theta$]
\addplot[blue, densely dotted, thick] table [x=theta2, y=AL, col sep=comma]{LR_CP_AL_rho_70.csv};
\addplot[red, loosely dashed, thick] table [x=theta2, y=AL, col sep=comma]{LRSC_CP_AL_rho_70.csv};
\addplot[teal, dashdotted,thick] table [x=theta2, y=AL, col sep=comma]{K18_CP_AL_rho_70.csv};
\addplot[violet, -, thick] table [x=theta2, y=AL, col sep=comma]{IICI_CP_AL_rho_70.csv};
\end{axis}
\draw (3,5.3) node {Panel C: Average Length};
\end{tikzpicture}
}
}
\end{center}
\vspace{-3mm}
\caption{Comparing the LRCI, SCLRCI, CLRCI, and IICI.\vspace{2mm}\\Notes: Panel A depicts the LRCI, SCLRCI, CLRCI, and IICI for $e'_1\theta$ when $e'_1\widehat\theta=0$ as functions of $e'_2\widehat\theta$, with $g(\theta)=e'_2\theta$ and $\text{corr}(e'_1\widehat\theta,e'_2\widehat\theta)=0.7$. 
The dotted blue lines depict the upper and lower bounds for the LRCI. 
The dashed red lines depict the upper and lower bounds for the SCLRCI. 
The dash-dotted green lines depict the upper and lower bounds for the CLRCI. 
The solid purple lines depict the upper and lower bounds for the IICI. 
Panels B and C depict the coverage probabilities and average lengths of the LRCI, SCLRCI, CLRCI, and IICI as functions of $e'_2\theta$, calculated with $10^5$ simulations. 
}
\label{Comparison_rho_70_2}
\end{figure}

\textbf{(1) LRCI/SCLRCI.} A natural approach to define a CI is to start with the likelihood-ratio statistic for testing $H_0: e'_1\theta=r, e'_2\theta\le 0$ against the alternative $H_1: e'_1\theta\neq r, e'_2\theta\le 0$, where $r$ is a hypothesized value of $e'_1\theta$. 
The LRCI is the set of values of $r$ for which the likelihood-ratio test fails to reject using the $1-\alpha$ quantile of the chi-squared distribution with one degree of freedom as the critical value. 
The LRCI is an attractive CI because it is equal to an upper-level set of the likelihood. 
The LRCI also reduces to the UCI when the inequality is sufficiently slack. 

The problem with the LRCI is that the distribution of the likelihood-ratio statistic is not chi-squared when there is an inequality on the nuisance parameters; the LRCI has not been shown to be valid. 
Simulations suggest the under-coverage is positive but small. 
A size-corrected version of the LRCI, SCLRCI, can be defined by taking the critical value to be the $1-\alpha$ quantile of the distribution of the likelihood-ratio statistic when the inequality is binding, calculated with $10^5$ simulations.\footnote{The SCLRCI is related to the second step of the two-step quasi-likelihood-ratio test in \cite{FanShi2023}. In \cite{FanShi2023}, the first step constructs a confidence set for a local slackness parameter and the second step estimates quantiles of the asymptotic distribution of the QLR statistic. The SCLRCI is also related to the test in \cite{CNPR2022}, which uses shrinkage in the bootstrap draws to approximate the quantile of the asymptotic distribution of the likelihood-ratio statistic when the inequality is binding.} 

The LRCI and SCLRCI are depicted in Figure \ref{Comparison_rho_70_2} with dotted blue lines and dashed red lines, respectively. 
They are practically indistinguishable because the size-correction is very small, although the SCLRCI is slightly longer. 
Compared to the IICI, the LRCI is equal to the IICI for most values of $e'_2\widehat\theta$. 
It is only when $e'_2\widehat\theta\approx\pm\ddot c$ that they differ. 
The differences in coverage probability and average lengths are barely visible in Panels B and C. 

The IICI can be viewed as a more nuanced size-correction to the LRCI than the SCLRCI. 
(Since the LRCI is only slightly invalid, only a minor adjustment should be needed.) 
The SCLRCI uses a different critical value that does not reduce to the UCI when the inequality is sufficiently slack---a blunt way to guarantee validity. 
It distorts the LRCI for all values of the slackness and ruins the ``never longer'' property of the LRCI. 
In contrast, the IICI modifies the LRCI only for values of $e'_2\hat\theta\approx\pm\ddot c$ and preserves the adaptation to the UCI/EICI when the inequality is estimated to be sufficiently slack/violated. 

\textbf{(2) CLRCI.} \cite{Ketz2018} recommends testing $H_0$ by simulating the conditional distribution of the likelihood-ratio statistic given a sufficient statistic for the slackness of the inequality. 
The CLRCI, implemented with $10^5$ simulations, is depicted in Figure \ref{Comparison_rho_70_2} with green dash-dotted lines. 
The CLRCI agrees with the UCI when the inequality is estimated to be sufficiently slack. 
As the inequality becomes less slack, the upper bound decreases until it is equal to $z_{1-\alpha}$. 
When the inequality is estimated to be violated, the lower bound on the CLRCI decreases. 
The CLRCI is usually (weakly) shorter than the UCI, but it can be longer. 
The CLR test for $H_0$ is essentially transitioning from a two-sided t-test to a one-sided t-test. 
Looking at Panel B, the coverage probability is always $95\%$, indicating that the CLRCI is similar. 
Looking at Panel C, the CLRCI has shorter average length than the IICI for some values of $e'_2\theta$ and longer average length for other values. 

\begin{figure}
\begin{center}
\scalebox{0.9}{
\subfigure{
\begin{tikzpicture}
\pgfmathsetmacro{\cov}{0.7}
\pgfmathsetmacro\horizontallength{6}
\def\verticallength{4.5}
\def\verticalshift{-0.5}
\pgfmathsetmacro\horizontalshift{0}
\pgfmathsetmacro{\EIEleft}{-\cov*(-\horizontallength+\horizontalshift)}
\pgfmathsetmacro{\EIEright}{-\cov*(\horizontallength+\horizontalshift)}
\begin{axis}[width=\textwidth, height=0.75\textwidth, ymin=-5,ymax=4,ylabel=$e'_1\theta$,xmin=-6,xmax=6,xlabel=$e'_2\widehat\theta$,legend style={at={(axis cs:(-5.4,-3)},anchor=north west}]
\draw[-] (-\horizontallength+\horizontalshift,0)--(\horizontallength+\horizontalshift,0); 
\draw[-] (0,-\verticallength+\verticalshift)--(0,\verticallength+\verticalshift); 
\addplot[blue, densely dotted, thick] table [x=theta2, y=upper, col sep=comma]{AG_rho_70.csv};
\addplot[red, dashed,thick] table [x=theta2, y=upper, col sep=comma]{EMW_rho_70.csv};
\addplot[teal, dashdotted, thick] table [x=theta2, y=upper, col sep=comma]{KM_rho_70.csv};
\addplot[violet, -, thick] table [x=theta2, y=upper, col sep=comma]{IICI_rho_70.csv};
\addplot[blue, densely dotted, thick] table [x=theta2, y=lower, col sep=comma]{AG_rho_70.csv};
\addplot[red, dashed,thick] table [x=theta2, y=lower, col sep=comma]{EMW_rho_70.csv};
\addplot[teal, dashdotted, thick] table [x=theta2, y=lower, col sep=comma]{KM_rho_70.csv};
\addplot[violet, -, thick] table [x=theta2, y=lower, col sep=comma]{IICI_rho_70.csv};
\addlegendentry{{\footnotesize IITCI}};
\addlegendentry{{\footnotesize EMWCI}};
\addlegendentry{{\footnotesize SSCI}};
\addlegendentry{{\footnotesize IICI}};
\end{axis}
\draw (\horizontallength+0.5,2*\verticallength+2*\verticalshift+3) node {Panel A: Confidence Intervals};
\end{tikzpicture}
}
}

\scalebox{0.9}{
\subfigure{
\begin{tikzpicture}
\begin{axis}[width=0.5\textwidth, height=0.4\textwidth, ymin=0.9,ymax=1,xmin=-5,xmax=0,xlabel=$e'_2\theta$]
\addplot[blue, densely dotted, thick] table [x=theta2, y=CP, col sep=comma]{AG_CP_AL_rho_70.csv};
\addplot[red, dashed,thick] table [x=theta2, y=CP, col sep=comma]{EMW_CP_AL_rho_70.csv};
\addplot[teal, dashdotted, thick] table [x=theta2, y=CP, col sep=comma]{KM_CP_AL_rho_70.csv};
\addplot[violet, -, thick] table [x=theta2, y=CP, col sep=comma]{IICI_CP_AL_rho_70.csv};
\end{axis}
\draw (3,5.3) node {Panel B: Coverage Probability};
\end{tikzpicture}
}
}
\scalebox{0.9}{
\subfigure{
\begin{tikzpicture}
\begin{axis}[width=0.5\textwidth, height=0.4\textwidth, ymin=3.2,ymax=4.2,xmin=-5,xmax=0,xlabel=$e'_2\theta$]
\addplot[blue, densely dotted, thick] table [x=theta2, y=AL, col sep=comma]{AG_CP_AL_rho_70.csv};
\addplot[red, dashed,thick] table [x=theta2, y=AL, col sep=comma]{EMW_CP_AL_rho_70.csv};
\addplot[teal, dashdotted, thick] table [x=theta2, y=AL, col sep=comma]{KM_CP_AL_rho_70.csv};
\addplot[violet, -, thick] table [x=theta2, y=AL, col sep=comma]{IICI_CP_AL_rho_70.csv};
\end{axis}
\draw (3,5.3) node {Panel C: Average Length};
\end{tikzpicture}
}
}
\end{center}
\vspace{-3mm}
\caption{Comparing the IITCI, EMWCI, SSCI, and IICI.\vspace{2mm}\\Notes: Panel A depicts the IITCI, EMWCI, SSCI, and IICI for $e'_1\theta$ when $e'_1\widehat\theta=0$ as functions of $e'_2\widehat\theta$, with $g(\theta)=e'_2\theta$ and $\text{corr}(e'_1\widehat\theta,e'_2\widehat\theta)=0.7$. 
The dotted blue lines depict the upper and lower bounds for the IITCI. 
The dashed red lines depict the upper and lower bounds for the EMWCI. 
The dash-dotted green lines depict the upper and lower bounds for the SSCI. 
The solid purple lines depict the upper and lower bounds for the IICI. 
Panels B and C depict the coverage probabilities and average lengths of the IITCI, EMWCI, SSCI, and IICI as functions of $e'_2\theta$, calculated with $10^5$ simulations.
}
\label{Comparison_rho_70_3}
\end{figure}

\textbf{(3) IITCI.} Another CI is based on the inequality-imposed estimator (IIE), which is defined by 
\begin{equation}
\text{IIE}=\begin{cases}
\widehat \theta&\text{ if }g(\widehat\theta)\le 0\\
\ddot\theta&\text{ if }g(\widehat\theta)>0
\end{cases}. 
\end{equation}
The inequality-imposed t-statistic (IIT) is formed using the IIE and the usual standard error ($\hat s$). 
The $\text{IITCI}=[\text{IIE}-\hat s z_{1-\alpha/2},\text{IIE}+\hat s z_{1-\alpha/2}]$ is depicted in Figure \ref{Comparison_rho_70_3} with blue dotted lines. 
The IITCI is equal to the UCI when the inequality is estimated to be slack. 
When the inequality is estimated to be violated, the IITCI is a translation of the UCI. 
Thus, the IITCI is always the same length as the UCI (also visible in Panel C). 
\cite{AndrewsGuggenberger2010ET} show using simulations that the maximum null rejection probability of the t-test using the IIT is $5\%$. 
This fact can be seen in Panel B because the coverage is monotonically increasing in $e'_2\theta$, so the minimum coverage is achieved in the limit when the inequality is infinitely slack. 

\textbf{(4) EMWCI.} \cite{ElliottMullerWatson2015} recommend a general strategy for calculating nearly weighted average power (WAP) optimal tests for a given hypothesis by specifying a WAP criterion, discretizing the null parameter space, and solving numerically for a distribution over the discretized null parameter space. 
They demonstrate the general strategy in this setting---testing a hypothesis when a nuisance parameter is subject to an inequality. 
Figure \ref{Comparison_rho_70_3} depicts the EMWCI with red dashed lines.\footnote{We use the switching version of the EMWCI, where the test is equal to the usual two-sided t-test when the inequality is estimated to be sufficiently slack.} 

In Panel A, the EMWCI is approximately the same as the UCI when the inequality is very slack. 
As the slackness decreases, the upper bound decreases slightly (although not monotonically). 
As the inequality is estimated to be more violated, the EMWCI approaches the IITCI. 
In Panel B, the coverage probabilities of the EMWCI and IICI are similar. 
In Panel C, there is a range of slackness values for which the EMWCI is shorter on average than the IICI, with the IICI shorter outside this range. 

The WAP of the IICI is $53.1\%$ with $2.5\times 10^5$ simulations, calculated with the WAP function specified in \cite{ElliottMullerWatson2015}.\footnote{This WAP is for the test that rejects when the CI does not include zero. Following \cite{ElliottMullerWatson2015}, the number of simulations is chosen so the simulation standard error is $0.1\%$.} 
This is the same as the WAP for the EMWCI reported in \cite{ElliottMullerWatson2015} and is within $\epsilon=0.5\%$ of their reported power bound of $53.5\%$. 
Thus, the IICI is nearly optimal for their WAP function. 

\textbf{(5) SSCI.} \cite{KetzMcCloskey2023} recommend a simple and short CI (SSCI). 
The SSCI is calculated by widening the UCI by a small amount. 
Then, an upper/lower bound makes the CI shorter. 
A constant is added to the bound to ensure the SSCI is valid. 
The SSCI is designed to trade off a shorter CI when the inequality is binding with a longer CI when the inequality is slack. 
Thus, the researcher should have a reasonable expectation that the inequality is binding or close to binding before implementing the SSCI. 
Figure \ref{Comparison_rho_70_3} depicts the SSCI with dash-dotted green lines. 
The SSCI is longer than the UCI when the inequality is sufficiently slack. 
When the inequality is violated, then the SSCI can be very short and even empty. 
In Panel B, the SSCI is conservative when the inequality is sufficiently slack. 
In Panel C, the average length of the SSCI is longer than the IICI over this range. 

\section{Empirical Applications}
\label{Applications}

\subsection{Cognitive Behavioral Therapy for Crime and Violence}
\label{Application1}

\cite{BlattmanJamisonSheridan2017} analyze an experiment designed to evaluate the effects of cognitive behavioral therapy on crime and violence. 
The experiment selected a sample of poor young men in Liberia that are high risk for crime and violence. 
The participants were randomized into one or both (or neither) of two treatments. 
The first treatment was eight weeks of cognitive behavioral therapy designed to formulate a law-abiding identity, encourage forward-looking reasoning and planning, and practice acceptable social behaviors. 
The second treatment was a cash grant of \$200 (USD, equivalent to about 3 months' wages). 
Participants were surveyed 2-5 weeks after treatment and 12-13 months after treatment. 

BJS use an index of ``antisocial activity'' that is calculated from survey questions on crime (drug selling, theft, arrests) and violence (fights, owning weapons, aggressive/hostile behavior, intimate partner abuse). 
The survey responses were validated using in-depth interviews of a subsample of participants. 
BJS estimate treatment effects\footnote{The treatment effects are for the ``intent-to-treat'' because not everyone completed the therapy sessions, although about two-thirds completed 80\% of the sessions.} by regressing the index of antisocial activity on treatment status, together with baseline characteristics and randomization-block fixed effects. 
The short-term effects are estimated using the surveys 2-5 weeks after treatment and the long-term effects are estimated using the surveys 12-13 months after treatment. 

Table \ref{LinearRegressionApplicationTable} reports estimates of the treatment effects, heteroskedasticity-robust standard errors, and UCIs (c.f. Table 2 in BJS). 
For the cash-only treatment, the short-term and long-term treatment effects are statistically insignificant. 
For the therapy-only treatment, the short-term treatment effect is large and statistically significant---a 0.25 standard deviation decrease in antisocial activity---while the long-term treatment effect is diminished and statistically insignificant. 
For both treatments combined, the short-term and long-term treatment effects are both large and statistically significant. 
BJS conclude that receiving the cash is equivalent to extending the effect of the therapy. 

\begin{table}[t]
\begin{center}
\scalebox{1}{
\begin{threeparttable}
\caption{Treatment Effects on Antisocial Behaviors}\label{LinearRegressionApplicationTable}
\begin{tabular}{lccc}
\hline\hline\vspace{-0.3cm}\\
&\multicolumn{3}{c}{2-5 Week Effects}\\
\cline{2-4}\vspace{-0.3cm}\\
&{Therapy Only}&{Cash Only}&{Both}\\
\hline\vspace{-0.3cm}\\
Estimate       & -0.249         & -0.079          & -0.308\\
Standard Error & (0.088)        & (0.091)        & (0.089)\\
UCI            & [-0.421,-0.076] & [-0.257,0.099] & [-0.481,-0.134]\\
IICI           & [-0.421,-0.076] & -              & [-0.481,-0.134]\\
IICI Length Ratio   & 1          & -              & 1\\
SSCI           & [-0.425,-0.072] & -              & [-0.485,-0.130]\\
SSCI Length Ratio   & 1.023          & -              & 1.023\\
\hline\\
&\multicolumn{3}{c}{12-13 Month Effects}\\
\cline{2-4}\vspace{-0.3cm}\\
&{Therapy Only}&{Cash Only}&{Both}\\
\hline\vspace{-0.3cm}\\
Estimate       & -0.083         & 0.132          & -0.247\\
Standard Error & (0.093)        & (0.097)        & (0.088)\\
UCI            & [-0.265,0.099] & \mbox{\hspace{5mm}}[-0.058,0.321]\mbox{\hspace{5mm}} & [-0.42,-0.074]\\
IICI           & \mbox{\hspace{5mm}}[-0.304,0.006]\mbox{\hspace{5mm}} & -              & \mbox{\hspace{5mm}}[-0.457,-0.169]\mbox{\hspace{5mm}}\\
IICI Length Ratio   & 0.852          & -              & 0.832\\
SSCI           & [-0.269,0.028] & -              & [-0.424,-0.147]\\
SSCI Length Ratio\mbox{\hspace{5mm}}   & 0.816          & -              & 0.799\\
\hline
\end{tabular}
{\small
\begin{tablenotes}
\item {\em Notes:} ``Estimate'' denotes the least squares estimator from a regression of antisocial behaviors on treatment statuses with a sample of $n=947$ young men. 
``Standard Error'' denotes the heteroskedasticity-robust standard error. 
``UCI'' denotes the usual confidence interval. 
``IICI'' denotes the inequality-imposed confidence interval. 
``SSCI'' denotes the short and simple confidence interval from \cite{KetzMcCloskey2023}. 
``Length Ratio'' denotes the length of the IICI or SSCI divided by the length of the UCI. 
\end{tablenotes}
}
\end{threeparttable}
}
\end{center}
\end{table}

\cite{KetzMcCloskey2023} point out that treatment effects in this application may have known sign. 
There are several reasons to think the cash-only treatment effect is nonpositive (does not increase antisocial activity). 
Survey data suggests participants spend little of the cash on temptation goods, including drugs, alcohol, gambling, or prostitution.\footnote{More generally, \cite{EvansPopova2017} conclude that cash transfers to poor individuals decreases expenditures on temptation goods.} 
This is validated by reported business investments/expenses, which accounts for more than half of the cash grant.\footnote{\cite{BlattmanFialaMartinez2014} and \cite{HaushoferShapiro2016} find that poor and unemployed individuals in East Africa have a high return to cash, possibly because they are credit constrained.} 
In addition, a pilot study found that the treatments were not harmful. 

Table \ref{LinearRegressionApplicationTable} reports CIs for the treatment effects assuming the cash-only treatment effect is nonpositive. 
We make the following remarks on Table \ref{LinearRegressionApplicationTable}. 

\textbf{Remarks.} (1) The IICIs and SSCIs are not calculated for the cash-only treatment. 
This is because they are only designed for inequalities on the nuisance parameters. 
A simple option for imposing the inequality on the UCI for the cash-only treatment effect is to intersect the UCI with $(-\infty,0]$, although \cite{MullerNorets2016} criticize the possibility that the resulting CI can be unreasonably short or empty. 

(2) For the short-term effects, the cash-only treatment effect is estimated to be sufficiently negative. 
Thus, the IICIs for the short-term effects are equal to the UCIs, demonstrating the adaptation of the IICI, while the SSCIs for the short-term effects are slightly longer. 

(3) For the long-term effects, the cash-only treatment effect is estimated to be positive. 
Under the assumption that the true cash-only treatment effect is nonpositive, this positive estimate is noise that can be removed by imposing the inequality. 
The IICIs and SSCIs for the long-term effects are shorter and more negative than the UCI. 
Thus, the IICIs and SSCIs indicate larger and longer-term treatment effects (less attenuation) of the therapy-only treatment and both treatments combined. 
This also suggests less importance for the cash follow-up in prolonging the effect of the therapy. 

(4) Comparing the IICIs and SSCIs for the long-term effects, note that the IICIs are shifted more negatively than the SSCIs. 
Also note that the SSCIs are shorter than the IICIs. 
In this case, the cash-only treatment effect is estimated to be sufficiently positive that the IICI is equal to the EICI. 
Thus, the SSCIs are shorter than the EICIs. 

\subsection{The Environmental Bias of Trade Policy}
\label{Application2}

\cite{Shapiro2021} analyzes the relationship between tariffs and NTBs and the carbon emissions of various industries. 
S21 investigates whether industries with higher carbon emissions, or ``dirty'' industries, receive an implicit subsidy from trade policy because they face lower tariffs and NTBs. 
S21 estimate a large and significant implicit subsidy for dirty industries. 

The data includes $n=2{,}021$ industry-country pairs. 
The dependent variables are tariffs or NTBs, measured as the average tax rate on imports of the output of a given industry into a given country.\footnote{NTBs are barriers to trade that are not tariffs, including price regulations, product standards, and quotas. S21 uses the ad-valorem equivalents of NTBs that are estimated in \cite{KeeNicitaOlarraga2009}.} 
Total carbon emissions includes direct carbon emissions, which are carbon emissions used to transform the input goods to the output good, and indirect carbon emissions, which are carbon emissions used to produce the input goods. 
Total carbon emissions are measured using a global input-output table that records, for each industry, how much of each input is used to produce \$1 of output. 
The fossil fuel inputs are converted to a measure of tons of carbon emissions. 
By inverting the input-output table, one can calculate a measure of total carbon emissions that includes direct and indirect emissions. 

As discussed in S21, this way of measuring total carbon emissions may include classical measurement error for two reasons. 
(1) The input-output table itself may include errors in the amount of input used to create the output. 
(2) There may be industry-specific prices paid for fossil fuels that is not accounted for in this measure. 
To deal with measurement error in the carbon emissions of an industry, S21 instrument for it using the direct emissions of that industry for the 10 other smallest countries. 
The smallest countries are more likely to take conditions in the rest of the world as given. 

The endogeneity of a regressor measured with error has the opposite sign as the coefficient. 
S21 discuss several reasons why one might expect the sign of the coefficient to be negative. 
Three reasons are: 
(a) The optimal tariff a large country places on imports from a small country is inversely proportional to the supply elasticity. 
This, combined with the fact that more differentiated industries may have lower elasticities and tend to be cleaner, would imply a negative coefficient. 
(b) Local politicians may impose lower tariffs or NTBs on dirty industries to try to relocate them to other countries to decrease local air pollution. 
(c) If firms lobby for lower tariffs on inputs and higher tariffs on outputs, and if final consumers are disorganized, then downstream industries will have higher tariffs. 
This, combined with the fact that upstream firms tend to be dirtier, would imply a negative coefficient. 
S21 ultimately conclude that the third reason best explains the implicit subsidy for dirty industries. 
(This conclusion overturns anecdotal evidence that dirty firms have outsized political influence.) 
For these reasons, it is reasonable to assume $\gamma$, the endogeneity of total carbon emissions, is nonnegative. 

\begin{table}[t]
\begin{center}
\scalebox{1}{
\begin{threeparttable}
\caption{Tariff and Nontariff Barriers and $\text{CO}_{\text{2}}$ Emissions Rates}\label{IVApplicationTable}
\begin{tabular}{lccccc}
\hline\hline\vspace{-0.3cm}\\
&\multicolumn{2}{c}{Unweighted} &&\multicolumn{2}{c}{Weighted by Trade Flow}\\
\cline{2-3}\cline{5-6}\vspace{-0.3cm}\\
&{Tariffs}&{NTBs}&&{Tariffs}&{NTBs}\\
\hline\vspace{-0.3cm}\\
Estimate       & -32.3         & -89.8          && -11.2&-75.7\\
S.E. & (8.4)        & (26.7)        && (5.4)&(29.4)\\
UCI            &\hspace{0.29cm}[-48.8,-15.8]\hspace{0.29cm}&\hspace{0.29cm}[-142.2,-37.4]\hspace{0.29cm}&& \hspace{0.29cm}[-21.7,-0.6]\hspace{0.29cm}&\hspace{0.29cm}[-133.3,-18.09]\hspace{0.29cm}\\
IICI          &\hspace{0.29cm}[-48.8,-15.8]\hspace{0.29cm}&\hspace{0.29cm}[-142.2,-39.2]\hspace{0.29cm}&& \hspace{0.29cm}[-21.7,-0.6]\hspace{0.29cm}&\hspace{0.29cm}[-133.3,-18.15]\hspace{0.29cm}\\
Length Ratio\hspace{0.29cm}   & 1          & 0.982              && 1& 0.999\\
$-\widehat\gamma$ ($\times 10^6$)       & -1.22 & -1.27 &  & -0.80 & -0.29 \\
$\widehat\gamma$ S.E. ($\times 10^6$) &(1.52)&(4.51)&&(0.54)&(2.97)\\
$\ddot c$ ($\times 10^6$)                    & 0.39 & 1.99 &   & 0.25 & 0.34 \\
\hline
\end{tabular}
{\small
\begin{tablenotes}
\item {\em Notes:} ``Estimate'' denotes the estimator from an IV regression of tariffs or NTBs on total carbon emissions and country fixed effects, using direct carbon emissions in small countries as an instrument. 
``S.E.'' denotes the standard error from that IV regression. 
``UCI'' denotes the usual confidence interval. 
``IICI'' denotes the inequality-imposed confidence interval assuming the endogeneity is nonnegative. 
``Length Ratio'' is calculated by taking the length of the IICI and dividing by the length of the UCI. 
``$-\widehat\gamma\hspace{0.5mm}$'' denotes the negative of the estimator of the endogeneity, estimated by GMM with the additional moment described in Example \ref{ExIV}. 
``$\widehat\gamma$ S.E.'' denotes the standard error of $\widehat\gamma$. 
``$\ddot c\hspace{0.5mm}$'' denotes the value of $\ddot c$ used in the formula for calculating the IICI. 
All standard errors are clustered at the industry level. 
\end{tablenotes}
}
\end{threeparttable}
}
\end{center}
\end{table}

We can estimate $\gamma$ jointly with the coefficients in the IV regression using GMM by adding a moment, as described in the last paragraph of Example \ref{ExIV}. 
S21 implement two versions of the IV regression, one with unweighted observations and one with observations weighted by trade flow. 
Table \ref{IVApplicationTable} reports results from these IV regressions. 
The first two rows of Table \ref{IVApplicationTable} report the coefficient estimates, together with their standard error, clustered at the industry level. 
The next three rows compare the UCI and the IICI. 
The final three rows report the estimates of the inequality and the value of $\ddot c$ used to calculate the IICI. 
We make three remarks on Table \ref{IVApplicationTable}. 

\textbf{Remarks.} (1) The first two rows in Table \ref{IVApplicationTable} replicate findings in Tables 2 and 3 in S21. 
The point estimates indicate a significant implicit subsidy for dirty industries through both tariffs and NTBs. 
As a point of comparison, the social cost of carbon emissions is usually estimated to be about \$40 per ton; see \cite{IWG2016}. 
The implicit subsidies from NTBs alone approximately double this amount. 

(2) Rows 3-5 in Table \ref{IVApplicationTable} compare the UCI and the IICI. 
For tariffs, the IICIs are equal to the UCIs. 
This again demonstrates the adaptation of the IICIs when the inequality is estimated to be sufficiently slack. 
For NTBs, the IICIs are shorter than the UCIs and exclude the smallest values of the implicit subsidy. 
Rows 6-8 in Table \ref{IVApplicationTable} explain why the IICIs are only slightly shorter than the UCIs (1.8\% in column 2 and 0.1\% in column 4). 
(The negative of $\widehat\gamma$ is reported in Row 6 in order to satisfy the setting in Section \ref{IICI_Definition}, where the inequality is assumed to be nonpositive.) 
In all columns, the inequality is not estimated to be violated, and it is only in columns 2 and 4 that the inequality is close enough to make a difference. 
The IICI gradually adjusts to the influence of the inequality, and therefore the IICI is only slightly shorter. 
This example demonstrates that the inequality need not be estimated to be violated in order for the IICI to be shorter than the UCI. 
It also demonstrates that the IICI is a subset of the UCI whenever $g(\widehat\theta)\le|\ddot c|$. 

(3) In the fourth column, the IICI is essentially equal to the UCI. 
This is also because $\ddot s$ is very close to $\hat s$, or, equivalently, the estimator of the endogeneity is approximately uncorrelated with the estimator of the coefficient. 
This demonstrates another adaptation property of the IICI: when $a'\widehat\theta$ is approximately uncorrelated with $e'_1\widehat\theta$, the IICI is approximately equal to the UCI. 
Even with this being true, unreported calculations show that the 99\% IICI for the fourth column excludes zero, while the 99\% UCI does not. 

\section{Conclusion}
\label{Conclusion}

This paper proposes a new CI, the IICI, for models with an inequality. 
The IICI has many attractive features, including being simple, adaptive, and never longer than the UCI. 
This paper proves finite-sample validity and asymptotic uniform validity and compares the IICI to alternatives available in the literature. 
This paper also demonstrates the IICI in two empirical applications. 

\section{Disclosure statement}\label{disclosure-statement}

The author reports there are no competing interests to declare. 

\section{Data Availability Statement}\label{data-availability-statement}

The data that support the findings of the empirical applications are openly available at \href{http://doi.org/10.1257/aer.20150503}{http://doi.org/10.1257/aer.20150503} 
and \href{http://doi.org/10.7910/DVN/CTUS2E}{http://doi.org/10.7910/DVN/CTUS2E}. 

\phantomsection\label{supplementary-material}
\bigskip

\begin{center}

{\large\bf SUPPLEMENTARY MATERIAL}

\end{center}

\begin{description}
\item[Title:] 
Online Appendix for ``A Simple and Adaptive Confidence Interval when Nuisance Parameters Satisfy an Inequality'' (pdf)
\end{description}

\bibliography{references.bib}

\end{document}